\documentclass[twocolumn,runningheads]{svjour2}
\smartqed  
\usepackage{graphicx}
\journalname{Astrophysics and Space Science}
\begin{document}

\title{Particle Acceleration at Multiple Internal Relativistic Shocks
\thanks{Paul Dempsey would like to thank IRCSET for their financial support.}
 }


\author{Paul Dempsey         \and
        Peter Duffy 
}


\institute{P. Dempsey \at
              UCD School of Mathematical Sciences, University College Dublin, Belfield, Dublin 4, Ireland. \\
              \email{paul.dempsey@ucd.ie}           
           \and
           P. Duffy \at
              UCD School of Mathematical Sciences, University College Dublin, Belfield, Dublin 4, Ireland.
}

\date{Received: 7 July 2006 / Accepted: 1 November 2006}

\maketitle

\begin{abstract}

Relativistic shocks provide an efficient method for high-energy particle acceleration in many astrophysical sources. Multiple shock systems are even more effective and of importance, for example, in the internal shock model of gamma-ray bursts. We investigate the reacceleration of pre-existing energetic particles at such relativistic internal shocks by the first order Fermi process of pitch angle scattering. We use a well established eigenfunction method to calculate the resulting spectra for infinitely thin shocks. Implications for GRBs and relativistic jets are discussed.

\keywords{Particle Acceleration \and Relativistic Shocks \and Internal Shocks}
\PACS{96.50.Pw \and 98.70.Sa}
\end{abstract}

\section{Introduction}
\label{intropdpd}
Observations of the relativistic jets associated with active galactic nuclei (AGN) show strong non-thermal emission at different sites within the jet structure and where the jet meets the intergalactic medium. While we cannot observe an outflow in gamma ray bursts (GRBs) a similar, but much more relativistic, jet structure is assumed with the prompt emission coming from the internal shocks while the external shock is responsible for the afterglow.

By analogy with the non-thermal emission from the non-relativistic shocks in supernova remnants, a first order Fermi process has been proposed by various authors as the mechanism responsible for the production of non-thermal particles at relativistic shock fronts. The semi-analytic eigenfunction method first introduced in the late 80's  \cite{Ref1pdpd,Ref2pdpd} allowed for calculation of the spectral index and anisotropy at shock front moving up to Lorentz factor 5 in the upstream rest frame. An improvement in
Kirk et al. \cite{Ref3pdpd} allowed a rapid computation of the spectral index of arbitrary shock speeds with the help of an analytic approximation.  The spectral index of particles accelerated at such shock fronts has been shown to tend to 4.22 for increasing Lorentz factors, in the hydrodynamic limit, where the compression ratio tends to 3. It has also been shown that the particle distribution is anisotropic at the shock front even for mildly relativistic flows.

Ignoring the injection mechanism, 
the jet produces a power law spectrum of electrons at a hydrodynamically dominated external shock via first order Fermi acceleration. This natural spectrum of the external shock, $s$, is calculated using the method of Kirk et al., of which we have given a brief summary. We then let these particle{{s}} escape downstream where they are accelerated again by an internal shock, which has natural index $q>s$. Adapting the work of Kirk et al., we are able to calculate the number of particles at the internal shock front compared to the number far upstream {at the external shock}. We can also measure the anisotropy at the shock and the number of particles far downstream of the internal shock.

{In Section 2 we summarise the semi-analytic eigenfunction method for determining the spectral index of a power law distribution produced at a shock and derive the distribution function far downstream of the internal shock. In Section 3 we describe the conservation of particle flux and use it to motivate a consistency check on our results, which are presented in Section 5.}

\section{An Eigenvalue Approach to The Transport Equation}
We restrict ourselves to the acceleration of particles with momenta well above the injection momentum but below any cut-off momentum where radiative losses change the shape of the spectrum. In the shock rest frame the particle transport equation, which holds separately upstream and downstream, can be written as 
\begin{equation}
\label{transEqn}
\Gamma\, (u+\mu)\frac{\partial f }{\partial z} =  \frac{\partial}{\partial \mu}D_{\mu\mu}\frac{\partial f}{\partial \mu}
\end{equation}
where {$u$ is the fluid velocity,} $\Gamma = (1-u^2)^{-1/2}$, $D_{\mu\mu}$ is the pitch-angle diffusion coefficient and $f$ is the particle phase-space distribution. We have to ensure that the distribution matches at the shock.
\begin{equation}
f^-(p_-,\mu_-,0)=f^+(p_+,\mu_+,0)
\end{equation}
where the plus (minus) sign denotes quantities upstream (downstream) of the shock.

We can expand $f$ as 
\begin{equation}
\label{EFeqn}
f(p,\mu,z)=\sum_{i=-\infty}^{\infty}b_i(p)Q_i(\mu)\exp\left(\frac{\Lambda_i z}{\Gamma}\right)
\end{equation}
where $(Q_i(\mu),\Lambda_i)$ are a{n} eigenfunction, eigenvalue pair satisfying 
\begin{equation}
\frac{d}{d \mu}D_{\mu\mu}\frac{d Q_i}{d \mu}=\Lambda_i(u+\mu)Q_i
\end{equation}
and the eigenvalues are ordered such that
\begin{eqnarray*}
\Lambda_{-i-1}<\Lambda_{-i}<\Lambda_{0}=0<\Lambda_{i}<\Lambda_{i+1} \qquad \forall i>0 
\end{eqnarray*}
To find a bounded solution with no particles far upstream we set $b_i(p)=a_ip^{-s}$ and the upstream solution becomes
\begin{equation}
f^-=\sum_{i>0}a_i^-p^{-s}_-Q_i^-(\mu_-)\exp\left(\frac{\Lambda_i^- z}{\Gamma}\right)
\end{equation}
with normalisation $a_1^-=1$, while the downstream can be written as
\begin{equation}
f^+=\sum_{i\le 0}a_i^+p_+^{-s}Q_i^+(\mu_+)\exp\left(\frac{\Lambda_i^+ z}{\Gamma}\right)
\end{equation}
then the matching condition can be reduced to 
\begin{equation}
\sum_{i>0}a_i^-p^{-s}_-Q_i^-(\mu_-)=\sum_{i\le 0}a_i^+p_+^{-s}Q_i^+(\mu_+)
\end{equation}
Using the orthogonality of the eigenfunctions we can then solve for $s$ and the $a_i$'s as in Kirk et al. In the limit of non-relativistic this gives the usual $s=4$ result with negligible anisotropy, while for highly relativistic shocks we reach $s=4.22$ with considerable anisotropy, approximately described in the shock rest frame as 
\begin{eqnarray*} 
{f(p_{\rm sh},\mu_{\rm sh},0)}\propto p_{\rm sh}^{-s}(1-\mu_{\rm sh}u_-)^{-s}\exp\left(-\frac{(1+\mu_{\rm sh})}{(1-u_-\mu_{\rm sh})}\right)
\end{eqnarray*}
Now suppose far upstream we have a known particle distribution $g(p_-)$. Then our matching condition for bounded solutions becomes
\begin{equation}
g(p_-)+\sum_{i>0} b_i^-(p_-)Q_i^-(\mu_-)=\sum_{i\le 0}b_i^+(p_+)Q_i^+(\mu_+)
\end{equation}
If our far upstream distribution is a power law $ g(p_-) = p_-^{-q} $ we can find solutions with $b_{i}^{\pm}(p_{\pm})=a_{i}^\pm p_{\pm}^{-q}$.

Using the Lorentz transformation $p_+= \Gamma_{\rm rel}p_-(1+u_{\rm rel}\mu_{-})$ where $u_{\rm rel} =(u_- - u_+)/(1-u_-u_+)$ the matching condition reduces to 
\begin{eqnarray}
(1+u_{\rm rel}\mu_-)^q +\sum_{i>0} a_i^-(1+u_{\rm rel}\mu_-)^qQ_i^-(\mu_-)\nonumber\\
=\sum_{i\le 0}a_i^+\Gamma_{\rm rel}^{-q}Q_i^+(\mu_+)
\end{eqnarray}
Then multiply by $(u_{+}+\mu_{+})Q^{+}_{j}(\mu_+)$, $j \ge 1$, and integrate over $\mu_{+}$ to get 
\begin{eqnarray}
\sum_{i>0}a_{i}^-W_{i,j} =-W_{0,j}/Q^{-}_{0}
\end{eqnarray}
where 
\begin{equation}
W_{i,j}=\int_{-1}^{1}(1+u_{\rm rel}\mu_-)^{q}(u_{+}+\mu_{+})Q^{+}_{j}(\mu_+)Q_{i}^-(\mu_-)d\mu_+
\end{equation}
so in matrix form ${\bf a} = {\bf W}^{-1}{\bf b}$ where $b_j=-W_{0,j}/Q^{-}_{0}$.

We now have the solution at the shock in the upstream rest frame. The isotropic distribution far downstream is thus
\begin{eqnarray}
f_d(p_+)&=&\frac{1}{2u_+} \int_{-1}^{1} f^-(p_-,\mu_-,z=0)(u_+ +\mu_+)\;d\mu_+
\end{eqnarray}

\section{Conservation of Particle Flux}
\label{sec:2pdpd}

In Heavens \& Drury \cite{Ref1pdpd} a consistency check was introduced to ensure conservation of particle flux in phase space. It can be explained as follows: consider all particles (upstream and downstream) with momentum up to $p_+$ as measured in the downstream rest frame. In the upstream rest frame this is all particles up to $p_-=\Gamma_{\rm rel}p_+(1+u_{\rm rel})$. The amount of particles in this range can only change, in the steady state, by, injection at momenta lower than $p_+$, flux from far upstream, flux lost far downstream and by particles accelerated at the shock by crossing from downstream to up and returning with momentum $p_+^*=p_+(1+u_{\rm rel})/(1-u_{\rm rel}\mu_+)$. This can be written as
\begin{eqnarray}
&\Phi& + \Gamma_-\int_{-1}^1\int_0^{p_-}(u_-+\mu_-)f(p_-^{\prime},\mu_-,-\infty)2\pi p_-^{\prime 2} dp_-^{\prime}d\mu_- \nonumber  \\
&=&\Gamma_+\int_{-1}^1\int_0^{p_+}(u_++\mu_+)f(p_+^{\prime},\mu_+,\infty)2\pi p_+^{\prime 2} dp_+^{\prime}d\mu_+ \nonumber \\
&+&\Gamma_+\int_{-1}^1\int_{p_+}^{p_+^*}(u_++\mu_+)f(p_+^{\prime},\mu_+,0)2\pi p_+^{\prime 2} dp_+^{\prime}d\mu_+
\end{eqnarray}
where $\Phi$ is the integrated injection flux up to $p_+$. Given $f(p_-,\mu_-,-\infty)=p_-^{-q}$, we differentiate with respect to $p_+$ and arrive at
\begin{eqnarray}\label{conserveDeriv}
2\Gamma_-u_-p_+^{-q}(\Gamma_{\rm rel}(1+u_{\rm rel}))^{-q+3}\nonumber\\ -\Gamma_+\int_{-1}^1(u_++\mu_+)f(p_+,\mu_+,\infty)d\mu_+ \nonumber =\\
\Gamma_+\int_{-1}^1(u_++\mu_+)\left(f(p_+^{*},\mu_+,0)\frac{(1+u_{\rm rel})^3}{(1-u_{\rm rel}\mu_+)^3} \right.\nonumber\\ 
- f(p_+,\mu_+,0)\Big)d\mu_+
\end{eqnarray}
Assuming the solution isotropises far downstream, the obvious solution is $f(p_+,\mu_+,0)=p_+^{-q}g(\mu_+)$, $f(p_+,\mu_+,\infty)=g_0p_+^{-q}$ where $g_0=\int(u_++\mu_+)g(\mu_+)d\mu_+/2u_+$ and equation (\ref{conserveDeriv}) reduces to 
\begin{eqnarray}
2\Gamma_-u_-(\Gamma_{\rm rel}(1+u_{\rm rel}))^{3-q} \nonumber\\
=\Gamma_+\int_{-1}^1(u_++\mu_+)\left(g(\mu_+)\frac{(1+u_{\rm rel})^{3-q}}{(1-u_{\rm rel}\mu_+)^{3-q}} \right)d\mu_+ 
\end{eqnarray}
When $q=3$ we see that $g_0=R$, the proper compression ratio. For $q\neq3$ we use 
\begin{equation}
\mathcal{N}_c=\frac{\Gamma_+\int_{-1}^1(u_++\mu_+)g(\mu_+)(1-u_{\rm rel}\mu_+)^{q-3}d\mu_+}{2\Gamma_-u_-\Gamma_{\rm rel}^{3-q}}
\end{equation}
as a measure of the consistency of our results{, where $\mathcal{N}_c=1$ implies an exact solution} .

\section{Adiabatic Gains}
In the absence of a diffusion process capable of scattering downstream particles back upstream, particles will undergo an energy change due to the compression of the plasma. Since the position element of phase space, $d^3x$ is compressed by $R$, the momentum element, $d^3p$, is expanded by $R$. Hence the downstream momentum can be related to the upstream momentum by $p_{+}=R^{-1/3}p_{-}$. Thus if we have $f(p_-,-\infty)=p^{-q}$ then the adiabatic gain is
\begin{eqnarray}
 \frac{f(z=\infty)}{f(z=-\infty)} = R^{q/3}
\end{eqnarray}
This is often referred to as ``shock-drift'' acceleration but it is in fact only the simplest form of it (for a more detailed view of shock-drift see Begelman \& Kirk \cite{Ref5pdpd}). We use this as a benchmark as it is the least amplification a power law can under go at a shock.

\section{Results}
The results presented in this poster consider the background plasma to have perpendicular magnetised shocks with negligible upstream hydrodynamic pressure, although it should {be} noted that they are also relevent for hydrodynamical shocks where the upstream pressure is important{, resulting in relatively weak shocks. This would be the case if the internal shock's upstream was the downstream of an external shock}.  The equation of state is taken to be Juttner-Synge with the downstream quantities solved from their upstream counterparts as in Kirk \& Duffy \cite{Ref4pdpd}.

First we will define a magnetisation value
\begin{equation}
 \sigma=\frac{v_A^2}{1-v_A^2}
\end{equation}
where $v_A$ is the Alfv\'en speed. 

\begin{figure}
 \centering
\includegraphics[width=.85\columnwidth]{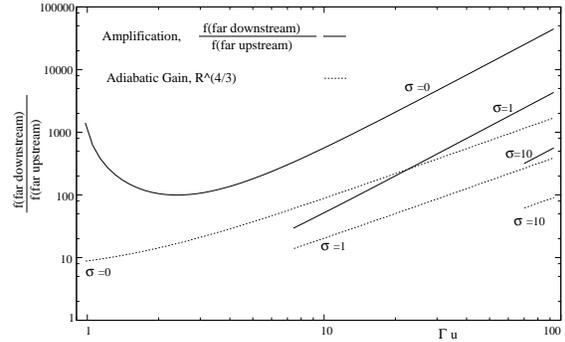}
 \caption{The effect {of} magnetisation on the amplification process with a $p^{-4}$ spectrum far upstream. As the magnetisation increases the compression ratio decrease, as does the amplification. {On the x-axis we have the 4-speed of the shock as measured in the upstream rest frame.}}
 \label{fig:1pdpd}       
 \end{figure}

 Figure~\ref{fig:1pdpd} illustrates the additional energy gain obtain{ed} from diffusive shock acceleration over shock drift acceleration for a population of cosmic rays with a power law distribution $p^{-4}$ advected into various relativistic shocks. Care must be taken when interpreting the results in the hydrodynamic case for low $\Gamma u$ as in this case $s\to q=4$ and the contribution of injected particles will be very important. However for large $\Gamma u$ the advected cosmic ray population with $p^{-4}$ will dominate over the $p^{-4.22}$ population originating from injection at the shock. It should be noted that the gain is more than one order of magnitude greater than that obtained via shock drift acceleration. So when ever the downstream plasma expands and the shock disappears the cosmic ray population will have been genuinely accelerated as the effects of adiabatic losses will not cancel the energy gain of the diffusive process as it would have the gain by shock drift.

\begin{figure}
 \centering
\includegraphics[width=.9\columnwidth]{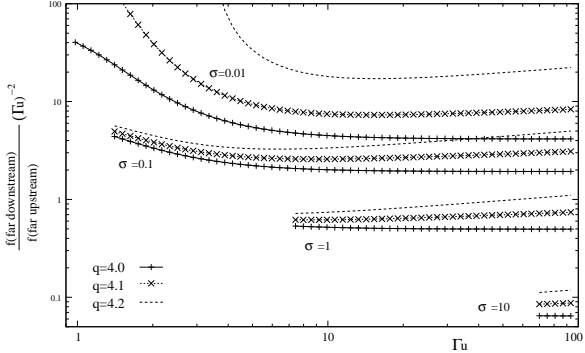}
 \caption{We measured the amplifications in the downstream rest frame for three power law distributions far upstream, $p^{-4}$, $p^{-4.1}$ and $p^{-4.2}$. On the x-axis we have the 4-speed of the shock as measured in the upstream rest frame. It seems that the energy gain from the diffusive process goes like $\Gamma_-^{q-2}$ for ultrarelativistic shocks.}
 \label{fig:2pdpd}       
 \end{figure}

Shocks with high $\sigma$ and a magnetic field perpendicular to the velocity have a reduced compression ratio, leading to an increase in their natural spectral index. However if there is a power law with a harder index upstream then we will still have amplification of it by the magnetised shock, and while it may be smaller than the amplification by its hydrodynamic equivalent, it is several orders greater than the non-relativistic prediction. Again the energy gain is greater than that achieved by shock drift acceleration, although we can see from the example of $\sigma=0.1$ that the energy gain difference is small for mildly relativistic shocks. However as the shock becomes more and more relativistic the difference between the energy gain from the two different processes increases. In fact, by looking at figure~\ref{fig:2pdpd} it seems that the energy gain from the diffusive process goes like $\Gamma_-^{q-2}$ for ultrarelativistic shocks, while the energy gain for shock drift is $R^{q/3}\simeq\Gamma_-^{q/3}$. Thus for ultra-relativistic shocks we have
\begin{eqnarray}
 \frac{\mbox{energy gain DSA}}{\mbox{energy gain shock drift}} \propto \Gamma_-^{2(q-3)/3}
\end{eqnarray}
This is the same scaling as noted in Begelman \& Kirk when they took a more detailed look at ``shock-drift'' acceleration in the ultrarelativistic limit.

\subsection{Accuracy and Consistency of the Method}
\begin{figure}
	\centering
	\includegraphics[width = .9\columnwidth]{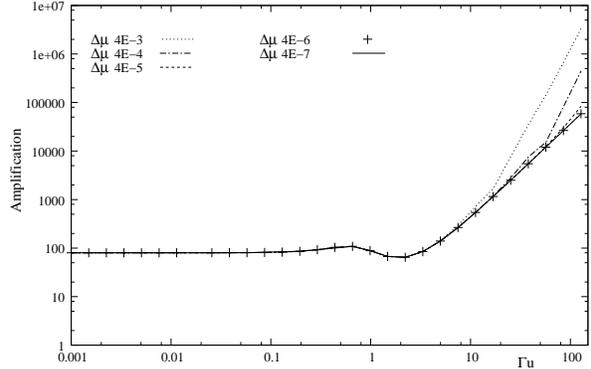}
	\caption{Convergence of the far downstream amplification, where upstream we have $p^{-3.95}$, as $\Delta\mu$ decreases. In our range of interest, $\Gamma <100$ the results converge completely by $\Delta\mu=4\times10^{-7}$.}
	\label{fig:3pdpd}
\end{figure}

While the method has been stated in a purely analytic manner, finding our eigenfunctions requires numerical integration, see Kirk et al. for details. As the left hand side of equation (\ref{EFeqn}) changes sign at $\mu=-u$ accurately sampling the functions between $\mu=-1$ and $\mu=-u$ is crucial, especially for highly relativistic flows. We chose a simple non-adaptive step size in $\mu$, $\Delta\mu$, and checked for the convergence of our solutions as we reduced $\Delta\mu$. The results shown in figure~\ref{fig:3pdpd} are for a power law with index $q=3.95$ advected into an unmagnetised strong shock satisfying the Juttner-Synge equation of state. 

Figure \ref{fig:3pdpd} shows how the convergence of the computed far downstream distribution with decreasing $\Delta\mu$. As expected the larger the Lorentz factor of the upstream medium in the shock rest frame the smaller we require $\Delta\mu$ to ensure an accurate result. Also the measure of consistency, $\mathcal{N}_c$, equals 1 for small enough $\Delta\mu$.

\section{Application to Jets with Internal Shocks}
If the shock is propagating into the ISM or IGM then we would expect $\sigma$ upstream of the shock to be small. If the shock is relativistic then it is expected that it is also a perpendicular shock. For perpendicular shocks it can be shown by a simple manipulation of the results presented in Kirk \& Duffy that
\begin{equation}
 \sigma_+=\sigma_-  \frac{u_-}{u_+}\left(\frac{1}{1+\sigma_-(1-u_-/u_+)}\right)
\end{equation}
where the velocities are measured in the shock rest frame. For small $\sigma_-$ at ultrarelativistic shocks ($v_-=1$) $v_+\approx1/3$ and 
\begin{equation}
 \sigma_+\approx 3  \sigma_-
\end{equation}
Thus the downstream of such a shock  {can} have a high $\sigma=\sigma_+^{e}${, unlike parallel shocks which reduce $\sigma$}. 
Due to the Lorentz invariance of $\sigma$ the upstream magnetisation value of this shock $\sigma_-^{i}$ will equal $\sigma_+^{e}$. This high magnetisation value reduces the compression ratio of the second shock allowing it to be a hydrodynamically weak, relativistic shock. Such shocks are not capable of producing hard power laws by themselves. However, assuming the external shock produces a hard power law, they can act as effective reacceleration sites throughout the jet, using the same pitch-angle diffusion mechanism responsible for production of hard power laws at strong shocks.



\end{document}